\documentclass[12pt,reqno]{amsart}

\usepackage{amssymb,amsthm,amsmath}
\usepackage[numbers,sort&compress]{natbib}
\usepackage{color}
\usepackage{graphicx}
\usepackage{tikz}
\usepackage{amssymb,amsthm,amsmath}
\usepackage[numbers,sort&compress]{natbib}
\usepackage{color}
\usepackage{graphicx}
\usepackage{tikz}
\usepackage{xparse}

\title[ ]{   Topics on Fermi varieties  of discrete periodic Schr\"odinger operators}
\author{Wencai Liu}
\address[ Wencai Liu]{ Department of Mathematics, Texas A\&M University, College Station, TX 77843-3368, USA}
 \email{liuwencai1226@gmail.com; wencail@tamu.edu}

\keywords{  Fermi variety, Bloch variety, analytic variety, algebraic variety, irreducibility, periodic Schr\"odinger  operator,  embedded eigenvalue, unique continuation,   rigidity theorem,  isospectrality, Floquet isospectrality,  Fermi isospectrality}

\subjclass[2010]{ 39A12 (primary); 14H10,  35J10, 47A75 (secondary)}

\theoremstyle{plain}
\newtheorem{theorem}{Theorem}[section]

\newtheorem{corollary}[theorem]{Corollary}
\newtheorem{lemma}[theorem]{Lemma}

\newtheorem{remark}{Remark}
\newcommand{\C}{\mathbb{C}}
\newcommand{\T}{\mathbb{T}}
\newcommand{\Z}{\mathbb{Z}}

\newcommand{\R}{\mathbb{R}}

\theoremstyle{plain}
\newtheorem{definition}{Definition}
\newtheorem{conjecture}{Conjecture}

\begin{document}
	\date{\today}
	
	\begin{abstract}
	  This is a survey  of recent progress on the   irreducibility of Fermi varieties,  rigidity results  and embedded eigenvalue problems of discrete periodic Schr\"odinger operators.

	\end{abstract}
	
	\maketitle 
	\section{Introduction}

	The geometry  of    Fermi varieties  plays critical roles in studying  many topics of    periodic Schr\"odinger 
	operators. In 	this article, we will first discuss the irreducibility of the  Fermi variety as an analytic set and  then   study  rigidity results (inverse problems)  and embedded eigenvalue problems of discrete periodic Schr\"odinger operators.
	We refer   readers to an earlier survey article ~\cite{kru} for  other  topics, techniques, and results 
	of both continuous and discrete periodic   Schr\"odinger  operators. 
	\subsection{Fermi, Bloch and Floquet varieties}
	Given $q_i\in \Z_+$, $i=1,2,\cdots,d$,
	let $\Gamma=q_1\Z\oplus q_2 \Z\oplus\cdots\oplus q_d\Z $.
	We say that a function $V: \Z^d\to \C$ is  $\Gamma$-periodic (or just periodic)  if 
	for any $\gamma\in \Gamma$ and $n\in\Z^d$,  $V(n+\gamma)=V(n)$. 
	
	For $n=(n_1,n_2,\cdots,n_d)\in\Z^d$, denote by 	$||n||_1=\sum_{i=1}^d |n_i|$.
	Let  $\Delta$ be the discrete Laplacian on lattices  $\Z^d$, namely
	\begin{equation*}
		(\Delta u)(n)=\sum_{n'\in\Z^d, ||n^\prime-n||_1=1}u(n^\prime).
	\end{equation*}
	Consider the  discrete    Schr\"{o}dinger operator on $\ell^2({\Z}^d)$ 
	\begin{equation} \label{h0}
		H_0=-\Delta +V .
	\end{equation}
	
	In this article, 
	we   always assume that $q_i$, $i=1,2,\cdots, d$, are relatively prime, $V$ is periodic and
	$H_0$ is the  discrete periodic Schr\"{o}dinger operator given by \eqref{h0}.
	
	Let $\{\textbf{e}_j\}$, $j=1,2,\cdots d$,  be the   standard basis in $\Z^d$:
	\begin{equation*}
		\textbf{e}_1 =(1,0,\cdots,0),\textbf{e}_2 =(0,1,0,\cdots,0),\cdots, \textbf{e}_{d}=(0,0,\cdots,0,1).
	\end{equation*}
	
	\begin{definition}
		The   { \it Bloch variety} $B(V)$ of  $-\Delta+V$ consists of all pairs $(k,\lambda )\in \C^{d+1}$ for which
		there exists a non-zero solution of the equation 
		\begin{equation} 
			(-\Delta u)(n)+V(n) u(n)=\lambda u(n) ,n\in\Z^d,  \label{spect_0}
		\end{equation}
		satisfying the so called Floquet-Bloch  boundary condition
		\begin{equation} 
			u(n+q_j\textbf{e}_j)=e^{2\pi i k_j}u(n),j=1,2,\cdots,d, \text{ and }  n\in \Z^d, \label{Fl}
		\end{equation}
		where $k=(k_1,k_2,\cdots,k_d)\in \C^d$.
		
	\end{definition}
	\begin{definition}
		Given  $\lambda\in \C$,
		the  Fermi surface (variety) $F_{\lambda}(V)$ is defined as the level set of the  Bloch variety:
		\begin{equation*}
			F_{\lambda}(V)=\{k: (k,\lambda)\in B(V)\}.
		\end{equation*}
	\end{definition}
	
	\begin{definition}
		Let  $\C^{\star}=\C\backslash \{0\}$ and $z=(z_1,z_2,\cdots,z_d)$. 
		The Floquet variety  is defined as 
		\begin{equation}\label{gff}
			\mathcal{F}_{\lambda}(V)=\{z \in (\C^\star)^d  :z_j=e^{2\pi i k_j}, j=1,2,\cdots,d,  k\in F_{\lambda}(V)\}.
		\end{equation}

	\end{definition}
	Clearly,
	\begin{equation*}
		F_{\lambda}(V)=\{k\in \C^d: (e^{2\pi ik_1}, e^{2\pi ik_2},\cdots, e^{2\pi ik_d})\in 	\mathcal{F}_{\lambda}(V)\}.
	\end{equation*}
	Namely,   $z\in (\C^\star)^d \in \mathcal{F}_{\lambda}(V)$
	if and only if  the equation
	\begin{equation}\label{flo}
		(-\Delta u) (n)+V(n)u(n)=\lambda u (n), n\in\Z^d
	\end{equation}
	with the boundary condition
	\begin{equation}\label{flo2}
		u(n+q_j\textbf{e}_j)=z_j u(n), j=1,2,\cdots, d, \text{ and } n\in \Z^d
	\end{equation}
	has  a   non-zero   solution.
	\subsection{Algebracity of the Floquet variety}
	
	Introduce a fundamental domain $W$ for $\Gamma$:
	\begin{equation*}
		W=\{n=(n_1,n_2,\cdots,n_d)\in\Z^d: 0\leq n_j\leq q_{j}-1, j=1,2,\cdots, d\}.
	\end{equation*}
	By writing out $-\Delta +V$  as acting on the  $Q=q_1q_2\cdots q_d$ dimensional space $\{u(n),n\in W\}$, 
	the eigen-equation \eqref{flo} and \eqref{flo2} (\eqref{spect_0} and \eqref{Fl})
	translates into the eigenvalue problem for a  $Q\times Q$ matrix $\mathcal{D}_V(z)$ ($D_V(k)$). 
	Let 
	$\mathcal{P}_V(z,\lambda)$ (${P}_V(k,\lambda)$) be the determinant of 
	$\mathcal{D}_V(z)-\lambda I$ (${D}_V(k)-\lambda I$).   
	
	It is easy to see that 
	\begin{equation}\label{g11}
		F_{\lambda}(V) =\{k\in\C^d: {P}_V(k,\lambda) =0\},
	\end{equation}
	and
	\begin{equation}\label{g111}
		\mathcal{F}_{\lambda}(V) =\{z\in(\C^\star)^d: \mathcal{P}_V(z,\lambda) =0\}.
	\end{equation}
	
	{\bf Example: $d=1$}
	\begin{equation*}
		D_V (k )= \begin{pmatrix}
			V(1)  &  -1  & 0&   \cdots & 0 & e^{-2\pi ik} \\
			-1 & V(2)  & -1 & \cdots   & 0& 0 \\
			0& -1 & V(3)&   \cdots& 0 & 0 \\
			\vdots & \vdots & \vdots & \ddots & \vdots & \vdots  \\
			0 & 0 & 0 & \cdots   & V(q_1-1)  & -1 \\
			e^{2\pi ik} & 0 & 0& \cdots  & -1 & V(q_1) \\
		\end{pmatrix}.
	\end{equation*}
	It is not difficult to see that $\mathcal{P}_V(z,\lambda) $ is a   polynomial  in variables $\lambda$ and 
	$z_1,z_1^{-1}, z_2,$\\$z_2^{-1}, \cdots,z_d,z_d^{-1}$.  See Lemma \ref{lesep} for a  more precise description. 
	In other words, $\mathcal{P}_V(z,\lambda) $ is a  Laurent polynomial  of $\lambda$ and 
	$z_1, z_2, \cdots,z_d.$
	Therefore,  the Floquet variety $\mathcal{F}_{\lambda}(V) $   is essentially an algebraic set.  
	
	\subsection{Discrete Floquet transform}
	Define the discrete Fourier transform $\hat{V}(l) $, $l\in {W}$ by 
	\begin{equation*}
		\hat{V}(l) =\frac{1}{{Q}}\sum_{ n\in {W} } V(n) \exp\left\{-2\pi i \left(\sum_{j=1}^d \frac{l_j n_j}{q_j} \right)\right\}.
	\end{equation*}
	For convenience, we extend $\hat{V}(l)$ to  a $\Gamma$-periodic function on $\Z^d$,  namely, for any $l\equiv m\mod \Gamma$,
	\begin{equation*}
		\hat{V}(l)=\hat{V}(m).
	\end{equation*}

	Let
	\begin{equation}\label{gtm}
		\tilde{\mathcal{D}}_V(z)= \tilde{\mathcal{D}}_V(z_1,z_2,\cdots,z_d)= \mathcal{D}_V(z_1^{q_1},z_2^{q_2},\cdots,z_d^{q_d}),
	\end{equation}
	and 
	\begin{equation}\label{gtp}
		\tilde{\mathcal{P}}_V(z,\lambda)=\det( \tilde{\mathcal{D}}_V(z,\lambda)-\lambda I)= \mathcal{P}_V(z_1^{q_1},z_2^{q_2},\cdots,z_d^{q_d},\lambda).
	\end{equation}
	Let $$\rho^j_{n_j}=e^{2\pi i \frac{n_j}{q_j}},$$
	where $0\leq n_j \leq q_j-1$, $j=1,2,\cdots,d$.

	By the standard discrete Floquet transform (e.g.,  \cite{liu1,ksurvey,kru}), one has
	\begin{lemma}\label{lesep} 
		Let $n=(n_1,n_2,\cdots,n_d) \in {W}$ and $n^\prime=(n_1^\prime,n_2^\prime,\cdots,n_d^\prime) \in {W}$. Then 
		$\tilde{\mathcal{D}}_V(z)$ is unitarily equivalent to 		
		$
		A+B_V,
		$
		where $A$ is a diagonal matrix with entries
		\begin{equation}\label{A}
			A(n;n^\prime)=-\left(\sum_{j=1}^d \left(\rho^j_{n_j}z_j+\frac{1}{\rho^j_{n_j} z_j} \right)\right) \delta_{n,n^{\prime}}
		\end{equation}
		and \begin{equation}\label{gb}
			B_V(n;n^\prime)=\hat{V} \left(n_1-n_1^\prime,n_2-n_2^\prime,\cdots, n_d-n_d^\prime\right).
		\end{equation}
		In particular,
		\begin{equation*}
			\tilde{\mathcal{P}}_V(z, \lambda) =\det(A+B_V-\lambda I).
		\end{equation*}

	\end{lemma}
	From Lemma \ref{lesep},  we have  that 
	$\mathcal{P}_V(z,\lambda) $ is a polynomial in  variables $\lambda$ and 
	$z_1,z_1^{-1},\cdots, z_d,z_d^{-1},$    with highest degree terms (up to a $\pm$ sign),
	$$\lambda^{Q}, z_1^{ \frac{Q}{q_1}}, z_1^{- \frac{Q}{q_1}}, z_2^{ \frac{Q}{q_2}},z_2^{- \frac{Q}{q_2}}, \cdots ,z_d^{ \frac{Q}{q_d}}, z_d^{- \frac{Q}{q_d}}.$$
	
	\subsection{Spectral bands}
	When $V$ is a real periodic function, $H_0$ is a self-adjoint operator and $D_V(k)$ is a Hermitian matrix  for any $k\in\R^d$.
	For each $k\in\R^d$,   $D_V(k)$ has $Q$ eigenvalues   and  order them in non-decreasing  order
	\begin{equation}\label{gspb}
		\lambda^1_V(k)\leq \lambda^2_V(k)\leq\cdots \leq \lambda^Q_V(k).
	\end{equation}
	We call   $\lambda^m_V(k)$ the $m$-th (spectral) band function, $m=1,2,\cdots, Q$.
	Let 
	\begin{align*}
		[a_m,b_m]= [\min_{k\in\R^d}\lambda^m_V (k),\max_{k\in\R^d}\lambda^m_V (k)], m=1,2,\cdots, Q. 
	\end{align*}
	Clearly, $a_m$ and $b_m$, $m=1,2,\cdots,Q$, depend on the potential $V$. We drop the dependence for simplicity. 
	$[a_m,b_m]$, $m=1,2,\cdots, Q$, are referred to as  spectral bands. 
	%
	%
	By the  standard Floquet theory \cite{ksurvey},  we have
	\begin{equation}\label{spg}
		\sigma_{ac}(H_0)=\sigma(H_0)=\cup_{m=1}^Q	[a_m,b_m], \sigma_{pp}(H_0)=\sigma_{sc}(H_0)=\emptyset.
	\end{equation}
	For the zero potential, one has that
	\begin{equation}\label{gfpg}
		\sigma_{ac}(-\Delta)=\sigma(-\Delta)=\cup_{m=1}^Q	[a_m,b_m]=[-2d,2d], \sigma_{pp}(-\Delta)=\sigma_{sc}(-\Delta)=\emptyset.
	\end{equation}

	\section{ Irreducibility of Fermi  and Bloch varieties}
	\subsection{  Irreducibility of Fermi  and Bloch varieties}
	In this section, we will discuss the irreducibility of both  Fermi and  Bloch   varieties as analytic sets. 
	
	\begin{definition}
		A subset   $\Omega\subset \C^k$ is called an analytic set if for any $x\in \Omega$, there is a neighborhood $U\subset \C^k$ of $x$, and analytic functions 
		$f_1,f_2,\cdots,f_p$ in $U$ such that
		\begin{equation*}
			\Omega\cap U=\{y\in U: f_1(y)=0,f_2(y)=0,\cdots,f_p(y)=0\}.
		\end{equation*}
	\end{definition}
	\begin{definition}
		An analytic set $ \Omega$ is said to be irreducible if it can not be represented as the union of two   non-empty proper  analytic subsets.  
	\end{definition}
	
	From \eqref{g11}, one can see that  the Fermi/Bloch variety  is a  principal analytic  set, namely, it is  determined by a single analytic function.   Since the Floquet-Bloch boundary condition is unchanged  under the shift: $k\to k+\Z^d$,  it is convenient to study Fermi and Bloch varieties
	modulo periodicity.  
	It is widely believed that the 
	Bloch/Fermi  variety (modulo periodicity)  is  always irreducible for periodic Schr\"odinger operators, which has been 
	formulated  as   conjectures:
	\begin{conjecture} ~\cite[Conjecture 5.17]{ksurvey} \label{conjb}
		The Bloch variety 
		$B(V)$  is irreducible  (modulo periodicity).
	\end{conjecture}
	\begin{conjecture} ~\cite[Conjecture 5.35]{ksurvey} ~\cite[Conjecture 12]{kvcpde20} \label{conj1dc}
		Let  $d\geq 2$.  Then 
		$F_{\lambda}(V)/\Z^d$  is irreducible, possibly except for  finitely 
		many $\lambda\in \C$.
	\end{conjecture}
	We remark that in Conjecture 1, 
	the irreducibility of  Bloch variety  modulo periodicity means 
	for any two irreducible components $\Omega_1$ and $\Omega_2$  of    $B(V)$, there exists $k\in\Z^d$ such that $\Omega_1=(k,0)+\Omega_2$.
	In Conjecture 2,  for fixed $\lambda$, 	$F_{\lambda}(V)/\Z^d$  is irreducible means 
	for any two irreducible components $\Omega_1$ and $\Omega_2$  of    $F_{\lambda}(V)$, there exists $k\in\Z^d$ such that $\Omega_1=k+\Omega_2$.
	
	Conjectures \ref{conjb} and \ref{conj1dc}   have  been mentioned in  many  articles ~\cite{batcmh92,bktcm91,ktcmh90,GKTBook,bat1,kv06cmp}.
	See  Conjecture 13 in ~\cite{kvcpde20} for a ``generic" version of Conjecture  \ref{conj1dc}.

	The irreducibility of Fermi and Bloch varieties  was   well understood for  $d=2,3$ about 30 years ago.
	When  $d=2$,    B{\"a}ttig  ~\cite{battig1988toroidal} proved that 	the Bloch variety 
	$B(V)$  is irreducible  (modulo periodicity). 
	In ~\cite{GKTBook},  Gieseker, Kn\"orrer and Trubowitz proved  that 
	$F_{\lambda}(V)/\Z^2$ is irreducible except for finitely many values of  $\lambda$.
	When   $d=3$,    B{\"a}ttig  ~\cite{batcmh92} proved that the Fermi variety 	$F_{\lambda}(V)/\Z^3$  is irreducible for every $\lambda\in\C$.  We refer readers to \cite{lslmp20,dksjmp20,flscmp} and references therein for more recent related results.
	
	In  ~\cite{GKTBook,bktcm91,bat1,batcmh92,battig1988toroidal},   proofs   heavily  depend on the construction of toroidal   and directional  compactifications of Fermi  and  Bloch varieties. 
	

	Recently, we introduced a new approach to study the Fermi variety and in particular   proved  both conjectures \cite{liu1}. For  any $d\geq 3$,  we proved that  the Fermi variety at every level  is irreducible  (modulo periodicity).     For $d=2$,    we proved that  the Fermi variety at every level except for the average of the potential  is irreducible  (modulo periodicity).   We also proved that   for any $d\geq 2$  the Bloch variety  is irreducible  (modulo periodicity).  
	
	Our approach focuses  on the study of the Laurent polynomial $\mathcal{P}(z,\lambda)$. This approach turns out to be quite robust and has already been  used  to  study the irreducibility of Bloch  varieties of discrete periodic operators on general lattices ~\cite{flm}.
	
	Before stating the main results in \cite{liu1}, some preparations are necessary.
	
	A    Laurent polynomial  of a  single term  is called monomial, i.e.,   $Cz_1^{a_1}z_2^{a_2}\cdots z_{k}^{a_k}$, where  $a_j\in\Z$, $j=1,2,\cdots,k$,
	and $C$ is a non-zero constant.
	\begin{definition}\label{de1}
		We say  that  a Laurent polynomial $h\left(z_{1}, z_{2},\cdots,z_k\right)$ is irreducible if it  can not be factorized non-trivially, that is, there are no   non-monomial Laurent polynomials $f\left(z_{1}, z_{2},\cdots,z_k\right)$  and $g\left(z_{1}, z_{2},\cdots,z_k\right)$ such that 
		$h=fg$.
	\end{definition}
	\begin{remark}
		When $h$ is a polynomial, the definition of irreducibility in Def.\ref{de1} is slightly different (allowing a difference of monomial) from the traditional definition (a polynomial is called irreducible if it can not be factorized into two non-constant polynomials).   For example, the polynomial $z^2+z$ is irreducible according to  Def.\ref{de1}.   
	\end{remark}

	Denote by $[V]$ the average of $V$:
	\begin{equation*}
		[V]=\frac{1}{Q}\sum_{n\in W} V(n).
	\end{equation*}
	
	Now we are ready to state the main results in 	\cite{liu1}.

	\begin{theorem}\label{gcf}
		\cite{liu1}
		Let $d\geq 3$.   Then for  any $\lambda\in \C$,  the Laurent polynomial $\mathcal{P}(z,\lambda) $ (as a function of $z$) is irreducible. 
		
	\end{theorem}

	\begin{theorem}\label{thm2}
		\cite{liu1}
		Let $d=2$.
		Then 	the Laurent polynomial   $ \mathcal{P}_V(z,\lambda)$ (as a function of $z$)  is irreducible for any $\lambda\in \C$ except for  $\lambda=[V] $. Moreover, if  $ \mathcal{P}(z,[V])$ is reducible,   
		$ \mathcal{P}(z,[V])$ has exactly two non-trivial  irreducible factors (count multiplicity).
	\end{theorem}

	\begin{theorem}\label{corbv1new}
		\cite{liu1}
		Let $d\geq 2$.  Then the  Laurent polynomial   $\mathcal{P}(z,\lambda) $ (as a function of both $z$  and $\lambda$) is irreducible. In particular,  the Bloch variety  $B(V)$ is irreducible  (modulo periodicity).
	\end{theorem}
	\begin{remark}\label{re2}
		\begin{enumerate}
			\item 	In 		\cite{liu1}, we actually proved that 
			when $ \mathcal{P}_V(z,[V])$ is reducible, $ \mathcal{P}_V(z,[V])=\mathcal{P}_{\bf{0}}(z,0) = ((-1)^{q_2}z_1^{q_2} +(-1)^{q_2}z_1^{-q_2}+(-1)^{q_1}z_2^{q_1}+(-1)^{q_1}z_2^{-q_1})$, where $\bf{0}$ is the  zero potential. 
			\item Theorem \ref{corbv1new} immediately follows  from Theorems \ref{gcf} and \ref{thm2} and some simple facts of the Laurent polynomial $\mathcal{P}(z,\lambda) $.
		\end{enumerate}
		
	\end{remark}
	\begin{corollary}\label{gcf1}
		\cite{liu1}
		For any $d\geq 3$, the  Fermi variety  $F_{\lambda}(V)/\Z^d$ is irreducible for any $\lambda\in \C$.
		For $d=2$,
		the  Fermi variety $F_{\lambda}(V)/\Z^2$  is irreducible for any $\lambda\in \C$ except  for $\lambda=[V] $. Moreover, if  $F_{[V]}(V)/\Z^2$ is reducible, it  has exactly two irreducible components.
		
	\end{corollary}
	
	\begin{corollary}\label{gcf2}
		\cite{liu1}
		For any $d\geq 3$, the  Floquet  variety  $\mathcal{F}_{\lambda}(V) $ is irreducible for any $\lambda\in \C$.
		For $d=2$,
		the  Floquet variety $\mathcal{F}_{\lambda}(V)$  is irreducible for any $\lambda\in \C$ except  for $\lambda=[V] $. Moreover, if  $\mathcal{F}_{[V]}(V)$ is reducible, it  has exactly two irreducible components.
		
	\end{corollary}

	\subsection{ Proof of Theorems \ref{gcf} and \ref{thm2} with  the zero potential}\label{S22}
	In order to better illustrate our ideas in 	\cite{liu1}, we first prove Theorems \ref{gcf} and \ref{thm2}  with the zero potential. 
	By Lemma \ref{lesep}, for the zero potential, 
	\begin{equation}\label{gtp1}
		\tilde{\mathcal{P}}_{\bf{0}}(z,\lambda)= \mathcal{P}_{\bf{0}}(z_1^{q_1},z_2^{q_2},\cdots,z_d^{q_d},\lambda)=(-1)^Q\prod_{n\in W}\left(\sum_{j=1}^d \left(\rho^j_{n_j}z_j+\frac{1}{\rho^j_{n_j} z_j} \right)+\lambda\right).
	\end{equation}
	
	Denote by  $\mu_{q_j}$  the multiplicative group of $q_j$ roots of unity, $j=1,2,\cdots, d$.   Let $\mu=\mu_{q_1}\times \mu_{q_2}\times \cdots \times \mu_{q_d}$.
	
	For any $\rho=(\rho^1,\rho^2,\cdots, \rho^d)\in \mu$, we  define a natural action  on $\C^d   $ 
	$$
	\rho \cdot\left(z_{1}, z_{2},\cdots, z_d\right)=\left(\rho^{1} z_{1}, \rho^{2} z_{2},\cdots,\rho^dz_d\right).
	$$

	\begin{proof}[\bf Proof of Theorem \ref{gcf} with the zero potential]
		Assume for some $\lambda\in\C$, $	{\mathcal{P}}_{\bf 0}(z,\lambda)=f(z)g(z)$ and both $f$ and $g$ are non-monomial  Laurent polynomials.
		Let
		\begin{equation}\label{g2}
			\tilde{f}(z)= {f}(z_1^{q_1},z_2^{q_2},\cdots,z_d^{q_d}),	\tilde{g}(z)= {g}(z_1^{q_1},z_2^{q_2},\cdots,z_d^{q_d}).
		\end{equation}
		By \eqref{gtp1} and \eqref{g2}, one has that 
		\begin{equation}\label{key1}
			\tilde{f}(z)	\tilde{g}(z)=(-1)^Q\prod_{n\in W}\left(\sum_{j=1}^d \left(\rho^j_{n_j}z_j+\frac{1}{\rho^j_{n_j} z_j} \right)+\lambda\right).
		\end{equation}
		When $d\geq 3$,  for any $n\in W$ and fixed $\lambda\in \C$, $\sum_{j=1}^d \left(\rho^j_{n_j}z_j+\frac{1}{\rho^j_{n_j} z_j} \right)+\lambda$ is irreducible as a function of $z$. 
		By the assumption that  $q_1,q_2,\cdots,q_d$ are relatively prime,  
		we have  that for any $n\in W,n'\in W$ with $n\neq n'$, 
		$\sum_{j=1}^d \left(\rho^j_{n_j}z_j+\frac{1}{\rho^j_{n_j} z_j} \right)+\lambda$ is not multiple of $ \sum_{j=1}^d \left(\rho^j_{n'_j}z_j+\frac{1}{\rho^j_{n'_j} z_j} \right)+\lambda $.
		Since both  $\tilde{f}(z) $  and $\tilde{g}(z)$  are  unchanged under the action $\mu$,
		we  have that  if $\tilde{f}(z) $  (or $\tilde{g}(z)$) has one factor $    \sum_{j=1}^d \left(\rho^j_{n_j}z_j+\frac{1}{\rho^j_{n_j} z_j} \right)+\lambda$,
		then  $\tilde{f} (z)$   ( or $\tilde{g}(z)$)  will have a factor 
		$$
		\prod_{ n\in W} \left(\sum_{j=1}^d \left(\rho^j_{n_j}z_j+\frac{1}{\rho^j_{n_j} z_j} \right)+\lambda\right).$$ This  contradicts  \eqref{key1}.
	\end{proof}
	\begin{proof}[\bf Proof of Theorem \ref{thm2} with the zero potential]
		When $\lambda\neq 0$,  it is easy to see that  for any $n\in W$, $\sum_{j=1}^2 \left(\rho^j_{n_j}z_j+\frac{1}{\rho^j_{n_j} z_j} \right)+\lambda$ is irreducible as a function of $z$. 
		By the similar   proof of Theorem \ref{gcf} with the zero potential, one has that 	${\mathcal{P}}_{\bf{0}}(z,\lambda)$ is irreducible.
		It suffices to prove the case that  $\lambda=0$. In this case,  by \eqref{gtp1},
		\begin{align}
			\tilde{\mathcal{P}}_{\bf{0}}(z,0)&=(-1)^Q  \prod_{n\in W}(\rho_{n_1}^1z_1+\rho_{n_2}^2z_2+\frac{1}{\rho_{n_1}^1z_1}+\frac{1}{\rho_{n_2}^2z_2})\nonumber\\
			&=(-1)^Q \left(\prod_{n\in W}(\rho_{n_1}^1z_1+\rho_{n_2}^2z_2)\right)  \left(\prod_{n\in W}\left(1+\frac{1}{\rho_{n_1}^1\rho_{n_2}^2z_1z_2}\right)\right) .\label{g3}
		\end{align}
		Since both $\prod_{n\in W}(\rho_{n_1}^1z_1+\rho_{n_2}^2z_2)$ and $\prod_{n\in W}\left(1+\frac{1}{\rho_{n_1}^1\rho_{n_2}^2z_1z_2}\right)$ are unchanged under $\mu$, one has that 
		there exist  a polynomial $f(z_1,z_2)$  and  a Laurent polynomial $g(z_1,z_2)$ such that
		\begin{equation*}
			f(z_1^{q_1},z_2^{q_2}) =\prod_{n\in W}(\rho_{n_1}^1z_1+\rho_{n_2}^2z_2)
		\end{equation*}
		and 
		
		\begin{equation*}
			g(z_1^{q_1},z_2^{q_2}) =\prod_{n\in W}\left(1+\frac{1}{\rho_{n_1}^1\rho_{n_2}^2z_1z_2}\right).
		\end{equation*}
		By the similar   proof of Theorem \ref{gcf}, both $f$ and $g$ are irreducible. 
		By \eqref{g3},  one has that $ {\mathcal{P}}_{\bf{0}}(z,0)=(-1)^Qf(z)g(z)$. This completes the proof.
	\end{proof}
	\subsection{ Ideas of the proof of Theorems \ref{gcf} and \ref{thm2} }
	In this subsection, we review the ideas of proof of Theorems \ref{gcf} and \ref{thm2}  in 	\cite{liu1}.
	Let $A_1=(0,0,\cdots, 0)$ (namely $z_1=z_2=\cdots=z_d=0$) and 
	$A_2=(0,0,\cdots, 0,\infty)$ (namely $z_1=z_2=\cdots=z_{d-1}=0$, $z_d^{-1}=0$).
	
	Assume  that $\mathcal{P}(z,\lambda)=\prod_{j=1}^mf_j(z)$, where $f_j$ is a (non-monomial) irreducible Laurent polynomial. 
	Let 
	\begin{equation}\label{g5}
		\tilde{f}_j(z)= {f}_j(z_1^{q_1},z_2^{q_2},\cdots,z_d^{q_d}), j=1,2,\cdots,m
	\end{equation}
	and hence
	\begin{equation}\label{g6}
		\tilde{\mathcal{P}}(z,\lambda)= \prod_{j=1}^m \tilde{f}_j(z).
	\end{equation}

	
	{\bf Step 1.}
	
	Let
	$$Z_{f_j}=	\{z\in (\C^{\star})^d: f_j(z)=0\} . $$
	Let $z_1=z_0^2$, $z_2=z_3=\cdots=z_{d-1}=z_0$ and $z_0\to 0$. 
	Solving  the equation $\tilde{\mathcal{P}}(z,\lambda)=0$  and by \eqref{A},  we   have that either 	   $z_{d}\to 0$ or $z_d^{-1}\to 0$.  This implies 
	that for any $j=1,2,\cdots,m$,  the closure (the   closure  is taken  in $(\C\cup \{\infty\})^d$)  of 
	$Z_{f_j}$   contains either 
	$A_1$  or $A_2$.
	
	{\bf Step 2.}
	
	We are going to define ``asymptotics" of $\mathcal{P}(z,\lambda)$/ $\tilde{\mathcal{P}}(z,\lambda)$  at $A_1=(0,0,\cdots, 0)$ and $A_2=(0,\cdots,0,\infty)$. 
	Let $ 	\tilde{h}_1(z)$ be the lowest (homogeneous)  degree component  of  $\tilde{\mathcal{P}}(z,\lambda)$, more precisely, 
	\begin{equation}\label{hom1}
		\tilde{h}_1(z)=	(-1)^Q \prod_{ n\in W} \left(\sum_{j=1}^d\frac{1}{\rho^j_{n_j}z_j}\right).
	\end{equation}
	Let $ 	\tilde{h}_2(z)$ be the lowest (homogeneous)  degree component  of $\tilde{\mathcal{P}}(z,\lambda)$ with respect to   variables $z_1,z_2,\cdots,z_{d-1},z_d^{-1}$, more precisely, 
	\begin{equation}\label{hom2}
		\tilde{h}_2(z)=(-1)^Q	\prod_{ n\in W} \left(\rho^d_{n_d} z_d+ \sum_{j=1}^{d-1}\frac{1}{\rho^j_{n_j}z_j}\right).
	\end{equation}
	Since both $	\tilde{h}_1(z)$ and $	\tilde{h}_2(z)$ are unchanged under the action $\mu$, there exist Laurent polynomials $h_1(z)$ and $h_2(z)$ such that 
	\begin{equation*}
		\tilde{h}_j(z)=h_j(z_1^{q_1},z_2^{q_2},\cdots, z_d^{q_d}), j=1,2.
	\end{equation*}
	Like previous arguments in Section \ref{S22},  it is not difficult to see  that both $h_1(z)$ and $h_2(z)$ are irreducible.
	
	{\bf Step 3.}
	
	Steps 1 and 2 allow us to conclude  that $\mathcal{P}(z,\lambda)$ has at most two non-trivial irreducible factors.
	Assume indeed, $\mathcal{P}(z,\lambda)$ has two  non-trivial irreducible factors, saying $f_1(z)$  and $f_2(z)$. Without loss of generality, assume
	the closure of $Z_{f_j}$  contains $A_j$, $j=1,2$.
	From Step 2, one can see that  $\tilde{f}_j(z)$  has   ``asymptotics" $\tilde{h}_j(z)$, $j=1,2$, namely
	up to monomials, 
	\begin{equation}\label{1g}
		\tilde{f}_1(z)=\tilde{h}_1(z)+\text{ higher degree terms of }  z_1, z_2,\cdots, z_{d-1},z_d
	\end{equation}
	and
	\begin{equation}\label{2g}
		\tilde{f}_2(z)=\tilde{h}_2(z)+\text{ higher degree terms of }  z_1, z_2,\cdots, z_{d-1}, z_d^{-1}.
	\end{equation}
	Finally, 	degree arguments  enable  us to  show that  two irreducible factors can only happen when $d=2$ and both higher degree terms in \eqref{1g} and \eqref{2g} vanish, namely, 
	up to monomials, 
	\begin{equation}\label{3g}
		\tilde{f}_j(z)=\tilde{h}_j(z), j=1,2.
	\end{equation}
	From \eqref{3g}, it is not difficult to  obtain that 
	$d=2$ and $\lambda=[V]$.

	We remark that the higher degree terms in \eqref{1g} and \eqref{2g} are with respect to different variables ($z_1,z_2,\cdots,z_{d-1}, z_d$ or $z_1,z_2,\cdots,z_{d-1},z_d^{-1}$)  and they may differ by monomials. Those obstacles make the  degree arguments  quite challenging.   Define the polynomial 		$\mathcal{P}_1(z,\lambda)=  (-1)^Qz_1^{\frac{Q}{q_1} }z_2^{\frac{Q}{q_2}}\cdots z_d^{\frac{Q}{q_d}}\mathcal{P}(z,\lambda)$.
	During the proof in 	\cite{liu1}, we need to  
	constantly work  between the polynomial $	\mathcal{P}_1(z,\lambda)	$ and the Laurent polynomial $\mathcal{P}(z,\lambda)$.

	\subsection{Open problems}
	
	Let $d=2$.
	Assume that the  Fermi variety $F_{\lambda}(V)$ is reducible at the average energy level $\lambda=[V]$. Does it imply the constancy of  the potential $V$ ?

	It is not difficult to see that    there exist   non-constant complex valued  functions $V$ such that the Fermi variety is reducible at the energy level $[V]$.  See ~\cite{liuprivate} for example.
	However,
	for real-valued potentials, people believe  the constant potential is the only case that 
	the Fermi variety $F_{\lambda}(V)$  is   reducible  at  some energy level, which has been formulated as a Conjecture  by Gieseker, Kn\"{o}rrer  and Trubowitz in the early 1990s  ~\cite{GKTBook}.
	
	\begin{conjecture} ~\cite[p.43]{GKTBook}\label{con4}
		Assume  that $V$  is  a non-constant real periodic potential on $\Z^2$. Then the  Fermi variety $F_{\lambda}(V)/\Z^2$  is irreducible for any $\lambda\in \C$.
	\end{conjecture}
	In a recent note,   we proved Conjecture \ref{con4}  for separable potentials ~\cite{liuprivate} (see next section for the definition of separable functions).
	
	Let  $\Delta_c$ be the continuous Laplacian on $L^2(\R^d)$:
	\begin{equation*}
		(\Delta_c u)(x)=\sum_{j=1}^d\frac{\partial^2 u}{\partial ^2x_j},x=(x_1,x_2,\cdots,x_d)\in\R^d.
	\end{equation*}
	Let $V_c:\T^d=\R^d/\Z^d\to\C^d$ be a periodic function. Like  discrete periodic Schr\"odinger operators, we 
	can  define the Fermi and Bloch varieties of  continuous periodic Schr\"odinger operators $-\Delta_c+V_c$. Both Fermi and Bloch varieties are (principal) analytic sets \cite{ksurvey}. 
	Conjectures \ref{conjb} and \ref{conj1dc}    are formulated for both continuous and discrete Schr\"odinger operators in \cite{ksurvey,kvcpde20}. 
	For continuous  periodic Schr\"odinger operators,  Kn\"orrer and Trubowitz proved  that  the Bloch variety    is irreducible  (modulo periodicity)
	when $d=2$  ~\cite{ktcmh90}. 
	
	
	Motivated by Theorems \ref{gcf} and \ref{thm2} and Conjecture \ref{con4} for  discrete periodic  Schr\"odinger  operators, we conjecture 
	that 
	
	\begin{conjecture} \label{conj4}
		Let $d\geq 3$.  Then  the Fermi    variety  of 
		$-\Delta_c+V_c$  is irreducible  (modulo periodicity)  for any $\lambda\in \C$.
	\end{conjecture}
	
	\begin{conjecture}  \label{conj5}
		Let $d= 2$.  Then  the Fermi    variety  of 
		$-\Delta_c+V_c$  is irreducible  (modulo periodicity)  for any $\lambda\in \C\backslash\{[V_c]\}$, where $[V_c]=\int_{\T^2}V_c(x)dx$.
	\end{conjecture}

	\begin{conjecture}  \label{conj6}
		Assume  that $V_c$  is  a non-constant real periodic potential on $\R^2$. Then  the Fermi    variety  of 
		$-\Delta_c+V_c$  is irreducible  (modulo periodicity)  for any $\lambda\in \C$.
	\end{conjecture}

	\section{Fermi isospectrality}
	\subsection{Isospectrality}
	In this section, we discuss  the  inverse problem of \eqref{spect_0} and \eqref{Fl} with  real periodic potentials.
	Two $\Gamma$-periodic potentials $V$ and $Y$ are called 
	Floquet isospectral if  
	\begin{equation}\label{gfi}
		\sigma(D_{V} (k))= \sigma(D_{Y} (k)), \text{ for any } k \in\R^d.
	\end{equation}

	We say a function $V$ on $\Z^d$ is $(d_1,d_2,\cdots,d_r)$ separable (or simply separable), where $\sum_{j=1}^r d_j= d$, if there exist functions 
	$V_j$  on $\Z^{d_j}$, $j=1,2,\cdots,r$,   
	such that  for any $(n_1,n_2,\cdots,n_d)\in\Z^d$,
	\begin{align}
		V(n_1,n_2,\cdots,n_d)=&V_1(n_1,\cdots, n_{d_1})+V_2(n_{d_1+1},n_{d_1+2},\cdots,n_{d_1+d_2})\nonumber\\
		&+\cdots+V_r(n_{d_1+d_2+\cdots +d_{r-1}+1},\cdots,n_{d_1+d_2+\cdots +d_r}).\label{g61}
	\end{align}
	
	We   say  $V:\Z^d\to \R$ is completely  separable if $V$ is $(1,1,\cdots, 1)$ separable. 
	When there is no ambiguity, we write   \eqref{g61} as $V=\bigoplus_{j=1}^r V_j$.
	
	Our interest  is first motivated by several questions asked   by  Eskin, Ralston and Trubowitz \cite{ERT84,ERTII}, and Gordon and Kappeler \cite{gki} : 
	\begin{enumerate}
		\item [Q'1.]   If  $Y$ and $V$ are Floquet isospectral,  and $Y$ is (completely) separable,  is $V$ (completely) separable?
		\item [Q'2.] Assume that both $Y=\bigoplus_{j=1}^dY_j$ and $V=\bigoplus_{j=1}^d V_j$   are completely separable. If $V$ and $Y$ are Floquet   isospectral,   are the one-dimensional potentials $V_j$ and $Y_j$ Floquet  isospectral (up to possible translations)?
	\end{enumerate}  
	While Q'1 and Q'2 were   formulated in the continuous case  \cite{ERT84,ERTII,gki,ksurvey}, the  same questions  apply to  the discrete case. 
	For both continuous and discrete periodic Schr\"odinger operators,  Q'1 and Q'2 have been partially answered   by  Eskin-Ralston-Trubowitz \cite{ERT84,ERTII}, Gordon-Kappeler \cite{gki} and Kappeler \cite{kapiii}.   
	We refer readers to
	~\cite{ksurvey,ERT84,MT76,kapiii,Kapi,Kapii,ERTII,gki,wa,gkii,gui90,eskin89} and references therein for precise descriptions of those  results  and    other related developments. 
	
	By \eqref{gspb}, one has that 
	\begin{equation}\label{g01}
		P_{V}(k)=\prod_{m=1}^Q (\lambda_V^m(k)-\lambda).
	\end{equation}
	Floquet isospectrality  of $V$ and $Y$ says that  for any $k\in\R^d$,  eigenvalues of eigen-equations \eqref{spect_0} and \eqref{Fl} with potentials 
	$V $ and $Y$ are the same.  This implies that Floquet isospectrality of $V$ and $Y$ is equivalent to   $F_{\lambda}(V)=F_{\lambda}(Y)$ for any $\lambda\in\C $ ($P_{V}(k)=P_{Y}(k)$ for any $k$; Bloch varieties of $V$ and $Y$ are the same).
	
	In   \cite{bktcm91},   B{\"a}ttig,  Kn\"orrer and Trubowitz studied  several rigidity  problems based on  the Fermi variety %
	when $d=3$.  In particular, they proved that 
	for continuous  periodic Schr\"odinger operators,  if at some energy level, Fermi varieties of a potential $V$ and the zero potential are the same, then $V$ is zero.
	This motivated  us to study the isospectrality based on Fermi varieties in 	\cite{liu2021fermi}, where    {\it Fermi isospectrality} was first introduced.

	
	\begin{definition}\label{fermiiso}	\cite{liu2021fermi}
		Let $V$ and $Y$ be two $\Gamma$-periodic functions. We say $V$ and $Y$ are  Fermi isospectral if 
		${F}_{\lambda_0} (V)={F}_{\lambda_0} (Y)$ for some $\lambda_0\in\C$. 
	\end{definition}

	Motivated by  Q'1 and Q'2, and the work of   B{\"a}ttig,  Kn\"orrer and Trubowitz  ~\cite{bktcm91},  we asked  two   questions in 	~\cite{liu2021fermi}:
	\begin{enumerate}
		\item [Q'3.]   If  $Y$ and $V$ are  {\it Fermi isospectral},  and $Y$ is (completely) separable,  is $V$ (completely) separable?
		\item [Q'4.] Assume that both $Y=\bigoplus_{j=1}^rY_j$ and $V=\bigoplus_{j=1}^r V_j$   are   separable. If $V$ and $Y$ are   {\it Fermi isospectral},   are the  lower dimensional potentials $V_j$ and $Y_j$ Floquet  isospectral?
	\end{enumerate}  
	In the same paper \cite{liu2021fermi},  we established  several rigidity theorems of  Fermi isospectrality  for  discrete periodic Schr\"odinger operators. In particular, we  answered  Q'3 and Q'4  affirmatively   for any dimension $d\geq 3$,  and thus answered  Q'1 and Q'2 as well.
	\subsection{Fermi isospectrality}

	%

	\begin{theorem}\label{thmmain2}
		\cite{liu2021fermi}
		Let  $d\geq 3$. Assume that $V$ and $Y$ are Fermi isospectral, and $Y$ is $(d_1,d_2,\cdots,d_r)$ separable, then
		$V$ is $(d_1,d_2,\cdots,d_r)$ separable.
	\end{theorem}
	
	As an immediate corollary, we have
	\begin{corollary}\label{coro1}
		\cite{liu2021fermi}
		Let  $d\geq 3$. Assume that $V$ and $Y$ are Fermi isospectral, and $Y$ is completely  separable, then
		$V$ is completely separable.
	\end{corollary}
	
	\begin{theorem}\label{thmmain3}
		\cite{liu2021fermi}
		Let  $d=d_1+d_2$ with $d_1\geq 2$ and $d_2\geq 1$. Assume that both $V$ and $Y$ are $(d_1,d_2)$ separable, namely, there exist $V_1, Y_1$ on $\Z^{d_1}$ and $V_2, Y_2$ on $\Z^{d_2}$ such that $V=V_1\oplus V_2$ and $Y=Y_1\oplus Y_2$.   Assume that $Y$ and $V$ are Fermi isospectral.   Then,  up to a constant, 
		$V_2$ and $Y_2$  are Floquet isospectral.
	\end{theorem}
	
	As a corollary of Theorems \ref{thmmain2} and \ref{thmmain3}, we have
	\begin{corollary}\label{coro2}
		\cite{liu2021fermi}
		Let  $d\geq 3$. Assume that $V$ and $Y$ are Fermi isospectral, and $Y=\bigoplus _{j=1}^d Y_j$ is completely  separable. Then $V=\bigoplus _{j=1}^d V_j$ is completely separable.   Moreover, up to a constant
		$Y_j$ and 	$V_j$, $j=1,2,\cdots,d$,   are Floquet isospectral.
	\end{corollary}
	\begin{theorem}\label{thmmain}
		\cite{liu2021fermi}
		Let $d\geq 3$.  
		Assume   that $V$  and   the zero potential  are Fermi isospectral. 
		Then
		$V$ is zero.
	\end{theorem}
	\begin{remark}\label{re00}
		For discrete periodic Schr\"odinger operators, Floquet isospectrality implies Fermi isospectrality,  so all results in this section  hold if we replace the assumption ``Fermi isospectrality  of $Y$ and $V$" with ``Floquet isospectrality of $Y$ and $V$".
	\end{remark}

	Let us talk about   ideas of the proof in \cite{liu2021fermi}.
	By the irreducibility results (Theorems \ref{gcf} and \ref{thm2}), we have that if $V$ and $Y$  are  Fermi isospectral, then $\mathcal{P}_V(z,\lambda)=\mathcal{P}_Y(z,\lambda)$ for some $\lambda\in \C$. Starting here, the proof in \cite{liu2021fermi}  focuses on the study of the Laurent polynomial $\mathcal{P}_V(z,\lambda)$ and $\mathcal{P}_Y(z,\lambda)$.  We first prove 
	\begin{theorem}\label{key11}
		\cite{liu2021fermi}
		Assume that  $V$ and $Y$ are Fermi isospectral. Then
		\begin{equation}\label{g21}
			[V]=[Y]
		\end{equation}
		and   for all possible $z\in\C^d$, 
		\begin{equation}\label{g55}
			\sum_{ n\in W, n^\prime \in W}
			\frac{|\hat{V}(n-n^\prime)|^2}{(\sum_{j=1}^d \rho^j_{n_j}z_j) (\sum_{j=1}^d \rho^j_{n^\prime_j}z_j) } \equiv \sum_{ n\in W, n^\prime \in W}
			\frac{|\hat{Y}(n-n^\prime)|^2}{(\sum_{j=1}^d \rho^j_{n_j}z_j) (\sum_{j=1}^d \rho^j_{n^\prime_j}z_j) }.
		\end{equation}
	\end{theorem}
	The following  proof depends on the study of linearly dependence/independence among hyperplanes $\sum_{j=1}^d \rho^j_{n_j}z_j=0$, $n\in W$ and  characterization  of separable functions in the   Fourier space. See 		\cite{liu2021fermi} for  details.
	\subsection{Open problems}
	Note that in Theorems \ref{thmmain2}, \ref{thmmain3} and   \ref{thmmain},  we assume $d\geq 3$. It is natural to ask the following questions:  
	\begin{enumerate}
		\item [Q1.]  Let $d=2$.  If  $Y$ and $V$ are  {\it Fermi isospectral},  and $Y$ is  separable,  is $V$  separable?
		\item [Q2.] 	Let  $d= 2$. Assume that both $V$ and $Y$ are   separable, namely, there exist $V_1, Y_1$ on $\Z$ and $V_2, Y_2$ on $\Z$ such that $V=V_1\oplus V_2$ and $Y=Y_1\oplus Y_2$.   Assume that $Y$ and $V$ are Fermi isospectral.   Is it true that up to a constant, 
		$V_1$ ($V_2$) and $Y_1$ ($Y_2$)  are Floquet isospectral?
		\item [Q3.] Let $d=2$.  
		Assume   that $V$  and   the zero potential  are Fermi isospectral. 
		Is it true that $V$ is zero?
	\end{enumerate}  
	By  Remark \ref{re2}, one can see that  the positive answer to Q3 implies Conjecture \ref{con4}. 
	From \cite{liuprivate}, we have that  the positive answer to Q1  implies   that to Q3 . 
	

	\section{Embedded eigenvalues}

	\subsection{Absence of embedded eigenvalues of perturbed periodic Schr\"odinger operators}
	
	Let us consider a perturbed  periodic Schr\"odinger operators:
	\begin{equation*}
		H=H_0+V=-\Delta+V+v,
	\end{equation*}
	where $V$ is periodic and $v$ is a decaying function.
	
	Recall that (see \eqref{spg})  the spectrum of $H_0=-\Delta+V$ has the  band structure and there are no eigenvalues.
	We are interested in  characterizing the perturbation $v$ such that $-\Delta+V+v$ has no eigenvalues embedded into the spectral bands, which is referred to as the embedded eigenvalue problem.
	
	For $d=1$, the  existence/absence  of  embedded eigenvalues   has been understood very well ~\cite{ld,krs,ns12,kn07,rof64,liucriteria,ld14,luk13}.
	Problems of the existence of embedded eigenvalues  in  dimension $d\geq 2$   are  a lot more complicated. The techniques of the  generalized Pr\"ufer transformation and oscillatory  integrals developed  for $d=1$  are not available.

	In ~\cite{kvcpde20}, Kuchment and Vainberg introduced  a    new approach to study the embedded  eigenvalue problem  for perturbed    periodic operators.
	It employs the analytic structure of the Fermi variety, unique continuation results, and techniques of several complex  variables theory.

	{\bf Condition 1:}  Given  $\lambda\in\bigcup(a_m,b_m)$, we say that  $\lambda$   satisfies Condition 1 if   any irreducible component of the   Fermi variety  at energy level $\lambda$ contains an open   analytic hypersurface   of dimension $d-1$ in $\R^d$. 
	
	\begin{remark}\label{re4}
		For $\lambda$ in the interior of a spectral band,	the irreducibility of the Fermi variety at energy  level $\lambda$ implies Condition 1 for the same $\lambda$ \cite{kv06cmp,kvcpde20,ksurvey}.
	\end{remark}

	\begin{theorem}\label{thm1kv}~\cite{kvcpde20}
		Let $d=2,3$,   $H_0=-\Delta_c+V_c$ and $H=-\Delta_c+V_c+v_c$.
		Assume that 	there exist constants $C>0$ and $\gamma>4/3$ such that
		\begin{equation*}
			|v_c(x)|\leq Ce^{-|x|^{\gamma}}, x\in\R^d.
		\end{equation*}
		Assume the Condition 1 for some  $\lambda\in \bigcup (a_m, b_m)$, where $\bigcup (a_m, b_m)$ is the spectral band of 
		$-\Delta_c+V_c$. 
		Then this $\lambda $ can not be an eigenvalue of $H=-\Delta_c+V_c+v_c$.
	\end{theorem}
	
	The restriction on $d=2,3$ and the critical exponent  $4/3$ arise  from a quantitative unique continuation result. 
	Suppose $u$ is a solution of 
	\begin{equation*}
		-\Delta_c u+\tilde{V}u=0 \text{ in } \R^d,
	\end{equation*}
	where $|\tilde{V}|\leq C$, $|u|\leq C$ and $u(0)=1$.
	From the unique continuation
	principle,  $u$ cannot vanish identically on any open set.
	The quantitative result states ~\cite{BK05}
	\begin{equation}\label{gcon43}
		\inf_{|x_0|=R}  \sup_{|x-x_0|\leq 1}|u(x)|  \geq  e^{-CR^{4 / 3} \log R}.
	\end{equation}

	A   similar   version  of  \eqref{gcon43} was established  in ~\cite{Mesh92} (also see Remark 2.6 in ~\cite{fhh}), namely, 
	there is  no non-trivial  solution of $(-\Delta_c +\tilde{V})u=0$ such that 
	\begin{equation}\label{gcon43w}
		|u(x)| \leq  e^{-C|x|^{4 / 3}} \text{ for any } C>0.
	\end{equation}

	For complex-valued  potentials $\tilde{V}$,
	the critical exponent  $ 4/3$  in  \eqref{gcon43} and \eqref{gcon43w} is optimal  by an example of  Meshkov ~\cite{Mesh92}. The Landis' conjecture states  that the critical exponent is 1 for real potentials.
	Recently, the Landis' conjecture was proved when $d=2$ by Logunov-Malinnikova-Nadirashvili-Nazarov \cite{logunov2020landis}. Following the proof of Theorem \ref{thm1kv} by Kuchment-Vainberg and the result of  Logunov-Malinnikova-Nadirashvili-Nazarov \cite{logunov2020landis}, one  could prove the following  result.
	\begin{theorem}\label{thm1kvnew} 
		Let $d=2$.
		Assume that $v_c$ is real and 	there exist constants $C>0$ and $\gamma>1$ such that
		\begin{equation*}
			|v_c(x)|\leq Ce^{-|x|^{\gamma}},x\in\R^2.
		\end{equation*}
		Assume the Condition 1 for some  $\lambda\in \bigcup (a_m, b_m)$, where $\bigcup (a_m, b_m)$ is the spectral band of 
		$-\Delta_c+V_c$. 
		Then this $\lambda $ can not be an eigenvalue of $H=-\Delta_c+V_c+v_c$.
	\end{theorem}

	The   unique  continuation principle for discrete Laplacians is well known not to  hold (see e.g., ~\cite{ku911,Jf07}).  
	In \cite{liu1}, we  observed that  a    weak  unique continuation result  is sufficient   for Kuchment-Vainberg's arguments 
	in  ~\cite{kvcpde20}.   
	Moreover,  the critical  component  can be  improved from ``$4/3$" to ``1". Therefore,
	we established the discrete version of Theorem \ref{thm1kv} for any dimension in \cite{liu1}.

	\begin{theorem}\label{thm1}
		\cite{liu1}
		Assume $V$ is a real-valued periodic function. 
		Let $d\geq 2$.
		Assume that  
		there exist constants $C>0$ and $\gamma>1$ such that the complex-valued function $v :\Z^d\to\C$  satisfies
		\begin{equation}\label{ggdecay}
			|v(n)|\leq Ce^{-|n|^{\gamma}},n\in\Z^d.
		\end{equation}
		Assume the Condition 1  for some $\lambda\in \bigcup_{m=1}^{Q}(a_m,b_m)$.
		Then this $\lambda $ can not be an eigenvalue of $H=-\Delta+V+v$.
	\end{theorem}
	The irreducibility result (established in Theorems \ref{gcf} and \ref{thm2},  and Remark \ref{re2})   and Remark \ref{re4} allow us to prove that 
	\begin{lemma}\cite{liu1}\label{lem4}
	Assume $V$ is a real-valued periodic function.   Then  the Condition 1  holds for every  $\lambda\in \bigcup_{m=1}^{Q}(a_m,b_m)$. 
	\end{lemma}
	By Theorem \ref{thm1} and Lemma \ref{lem4},  we have that 
	\begin{theorem}\label{thm11}
		\cite{liu1}
	Assume $V$ is a real-valued periodic function.
		If  
		there exist constants $C>0$ and $\gamma>1$ such that the complex-valued function $v :\Z^d\to\C$ satisfies 
		\begin{equation}\label{ggdecaynew}
			|v(n)|\leq Ce^{-|n|^{\gamma}},
		\end{equation}
		then   $H=-\Delta+V+v$ does not have any embedded eigenvalues, i.e., for any $\lambda \in \bigcup_{m=1}^{Q}(a_m,b_m)$, $\lambda$ is not an eigenvalue of $H$.
	\end{theorem}

	Assume that $V$ is zero.   We  can think that  $V$  is a  periodic function on $\Z^d$ with any  $q_1,q_2,\cdots,q_d$. 
	Recall \eqref{gfpg}. 
	
	\begin{lemma}~\cite[Lemmas 1.2 and 1.3]{hj18}\label{Enot0}
		Let $d\geq 2$.
		Then
		\begin{itemize}
			\item for any  $\lambda\in (-2d, 2d)\setminus \{0\}$,  $\lambda \in  (a_m,b_m)$ for some $1\leq m\leq Q$,
			\item  	if  at least one of $q_j$'s is odd, then $0\in (a_m,b_m)$ for some $1\leq m \leq Q$.
		\end{itemize}
	\end{lemma}
	
	For $d=2$, Lemma  \ref{Enot0}    was  also  proved in ~\cite{ef}.
	
	Theorem \ref{thm11} and Lemma \ref{Enot0} imply 
	\begin{corollary}\label{cor14}
		\cite{liu1}
		Assume that there exist some $C>0$ and $\gamma>1$ such that 
		\begin{equation}\label{g10}
			|v(n)|\leq Ce^{-|n|^{\gamma}}.
		\end{equation}
		Then 
		$\sigma_{p}(-\Delta+v)\cap(-2d,2d)=\emptyset$.
	\end{corollary}
	\begin{remark}\label{re5}
		\begin{enumerate}
			\item 	When $d=1$,  stronger results (replace \eqref{ggdecaynew} and \eqref{g10} with $v(n)=\frac{o(1)}{1+|n|}$ as $n\to\infty$) in Theorem \ref{thm11} and Corrollary   in \ref{cor14} have been proved  \cite{liucriteria,ld,rof64}. 
			\item Under a stronger  assumption that $v$ has compact support,   Isozaki and Morioka  proved that $\sigma_{ p}(-\Delta+v)\cap(-2d,2d)=\emptyset$ ~\cite{IM14}.
		\end{enumerate}

	\end{remark}
	\begin{remark}
		\begin{enumerate}
			\item For periodic (quantum) graph operators, the reducibility of  Fermi varieties  may occur.    (quantum) graph operators  may  lack      unique continuation results.  There are examples showing that   embedded eigenvalues (even with compactly supported   eigenfunctions) happen under compactly supported  perturbations ~\cite{kv06cmp,ku911,sjst,flscmp,scmp,kp07}. 
			\item For the continuous fourth order differential operator $D^4$ on $L^2(\R)$, under  a compactly supported   perturbation $v_c$, $D^4+v_c$ can have embedded eigenvalues ~\cite{pap}. 
			\item For  the continuous fourth order differential operator on periodic hexagonal lattices, the reducibility of Fermi varieties occurs \cite{MH21}.
		\end{enumerate}
	\end{remark}

	\subsection{Open problems}
	We list a few questions here.

	\begin{enumerate}
		\item[Q4.] Let   $v_c$ be a real function  on $\R^d$. 
		Assume that the Landis' conjecture holds and the Condition 1 holds  for some $\lambda\in \bigcup(a_m,b_m)$.   
		Assume that 
		there exist constants $C>0$ and $\gamma>1$ such that
		\begin{equation*}
			|v_c(x)|\leq Ce^{-|x|^{\gamma}}.
		\end{equation*}
		Then  following the argument of Kuchment-Vainberg \cite{kvcpde20}, $\lambda $ can not be an eigenvalue of $H=-\Delta_c+V_c+v_c$. 
		We ask whether or not  we could  find an alternative  proof which does not depend  on  Landis' conjecture  or Condition 1.
		\item [Q5.] 
		For continuous Schr\"odinger operators, 	when the periodic function $V_c=0$,  Kato \cite{kato} proved that $-\Delta_c+v_c$ does not have embedded eigenvalues if $\frac{o(1)}{1+|x|}$ as $x\to \infty$ (the actual result is quantitative).
		Motivated by  results of Kato  and one dimensional case (see (1) in Remark \ref{re5}),   we ask  is it true that $-\Delta+V+v$ ($-\Delta_c+V_c+v_c$) does not have any embedded eigenvalues if  $|v(n)|\leq \frac{C}{|n|^K},n\in\Z^d$ ($|v_c(x)|\leq \frac{C}{|x|^K},x\in\R^d$ ) for some $K,C>0$?
		We should mention  that for the  discrete case,  this is open even for $V=0$.
		\item [Q6.] Contruct properly decaying functions $v$  ($v_c$) such that $-\Delta+V+v$  ($-\Delta_c+V_c+v_c$) has  embedded eigenvalues.  Here ``properly decaying"  means exponentially decaying, algebraic decaying or something similar.  mean When $d=1$, there are many  examples of perturbed periodic Schr\"odinger  operators with (dense)embedded eigenvalues \cite{nabo,sim,ld,liucriteria,lj,liujfa,liumana,jnw,jnw1,jnw2}. The construction is quite challenging when $d\geq 2$ and the underlying periodic functions ($V$ or $V_c$) are non-zero.  	For the continuous case, when $V_c=0$, it reduces to an one dimensional problem   by choosing radial functions $v_c(x)=v_c(r)$, where $r=\sqrt{x_1^2+x_2^2+\cdots +x_d^2}$.
		For the discrete case,  when $V=0$, it can be done by  taking   separable functions $v(n)=v_1(n_1)+v_2(n_2)+\cdots +v_d(n_d)$.

	\end{enumerate}
	
	\section*{Acknowledgments}
	
	W. Liu was 
	supported by NSF  DMS-2000345 and DMS-2052572.

\end{document}